\documentclass[twocolumn,preprintnumbers,amsmath,amssymb,prx]{revtex4}
\usepackage[dvipdfmx]{graphicx}
\usepackage{graphicx}
\usepackage{dcolumn}
\usepackage{bm}
\usepackage{color}
\usepackage{rotating}
\usepackage[normalem]{ulem}

\begin{document}

\title{Charge neutral fermions and magnetic field driven instability in insulating YbIr$_3$Si$_7$}

\author{Y.\,Sato$^1$}
\author{S.\,Suetsugu$^1$}
\author{T.\,Tominaga$^1$}
\author{Y.\,Kasahara$^1$}
\author{S.\,Kasahara$^1$}
\author{T.\,Kobayashi$^1$}
\author{S.\,Kitagawa$^1$}
\author{K.\,Ishida$^1$}
\author{R.\,Peters$^1$}
\author{T.\,Shibauchi$^2$}
\author{A. H.\,Nevidomskyy$^3$}
\author{L. Qian$^{3,4}$}
\author{J. M. Moya$^{3,4}$}
\author{E.\,Morosan$^{3,4}$}
\author{Y.\,Matsuda$^1$}

\affiliation{$^1$ Department of Physics, Kyoto University, Kyoto 606-8502 Japan}
\affiliation{$^2$Department of Advanced Materials Science, University of Tokyo, Kashiwa, Chiba 277-8561, Japan}
\affiliation{$^3$ Department of Physics and Astronomy, Rice University, Houston, TX 77005 USA} 
\affiliation{$^4$ Department of Chemistry, Rice University, Houston, TX 77005 USA}


\pacs{}

\begin{abstract}
Materials where localized magnetic moments are coupled to itinerant electrons, the so-called Kondo lattice materials, provide a very rich backdrop for strong  electron correlations. They are known to realize many exotic phenomena, including unconventional superconductivity,  strange metals, and  correlated topological phases of matter. 
Here, we report what appears to be electron fractionalization in insulating Kondo lattice material YbIr$_3$Si$_7$, with emergent neutral excitations that carry heat but not electric current and contribute to metal-like specific heat.
We show that these neutral particles change their properties as the material undergoes a transformation between two antiferromagnetic phases in an applied magnetic field.
In the low-field AF-I phase, we find that the low temperature linear specific heat coefficient $\gamma$  and the residual linear term in the thermal conductivity $\kappa/T(T\rightarrow 0)$ are finite, demonstrating itinerant gapless excitations.  These results, along with a spectacular violation of the Wiedemann-Franz law,  directly indicate that YbIr$_3$Si$_7$ is a charge insulator but a thermal metal.   Nuclear magnetic resonance spectrum  reveals a spin-flop transition to a high field AF-II phase.  Near the transition field, $\gamma$ is significantly enhanced. Most surprisingly, inside the AF-II phase, $\kappa/T$ 	exhibits a sharp drop below $\sim300$\,mK,  indicating either opening of a tiny gap or a linearly vanishing density of states. 
This finding demonstrates a transition from a thermal metal into an insulator/semimetal driven by the spin-flop magnetic transition.  These results suggest that spin degrees of freedom directly couple to the neutral fermions, whose emergent Fermi surface undergoes a field-driven instability at low temperatures.
\end{abstract}

\maketitle

\section{Introduction}
Strong electron interactions often lead to the emergence of  many-body insulating ground states.   Recently, surprising properties  have  aroused considerable interest in  the research of the strongly correlated insulators, SmB$_6$ and YbB$_{12}$  with simple cubic crystal structures \cite{Takabatake}.  In these Kondo lattice compounds,  the band gap opens up at low temperatures due to the hybridization of localized $f$-electrons with conduction electrons \cite{Riseborough}.    In particular,  quantum oscillations (QOs) \cite{Li,Tan,Xiang, Hartstein, Xiang2021}, specific heat \cite{Hartstein,Sato,Terashima}, and thermal conductivity \cite{Sato,Hartstein,Boulanger,YXu} experiments have posed a significant paradox, revealing gapless excitations  in the bulk,  in apparent contradiction with the charge gap seen in transport measurements.  While the angular dependence of the QO frequencies suggests a three-dimensional (3D)  bulk Fermi surface in SmB$_6$ \cite{Tan} and YbB$_{12}$ \cite{Xiang}, both materials remain robustly insulating to very high magnetic fields.  (In SmB$_6$, a 2D Fermi surface has also been reported \cite{Li,Thomas}).   Various theoretical models of the QOs in these insulators have been proposed so far \cite{Knolle1,Knolle2,Chowdhury,Sodemann,Erten,Varma,Shen,Peters,Tada,Baskaran}.  Another striking aspect is a nonzero low temperature linear specific heat coefficient $\gamma\sim10$\,mJ/K$^2$mol for SmB$_6$ \cite{Hartstein} and $\sim4$\,mJ/K$^2$mol for  YbB$_{12}$ \cite{Sato,Terashima} in zero field.   As the specific heat is measured in the bulk insulating state,  these results indicate  the existence of  gapless and charge-neutral excitations in the bulk consistent with an emergent Fermi surface of neutral fermions.

However, there are distinct differences in the gapless excitations in these correlated  insulators.  In SmB$_6$, the QOs are observed only in the magnetization (de Haas-van Alphen, dHvA,  effect).  The dHvA oscillations strongly deviate below 1\,K from the Lifshitz-Kosevich  theory, which is based on  Fermi liquid theory \cite{Tan}.   In contrast, in YbB$_{12}$,  the QOs are observed not only in the magnetization, but also in the resistivity (Shubnikov-de Haas, SdH, effect) and  both dHvA and SdH oscillations obey the Lifshitz-Kosevich  theory down to 50\,mK \cite{Xiang}.   Moreover,  in YbB$_{12}$, a finite residual temperature-linear ($T$-linear)  term in the  thermal conductivity $\kappa_0/T\equiv \kappa/T(T\rightarrow 0)$ is observed, demonstrating  the presence of gapless and itinerant neutral fermions \cite{Sato}.  On the other hand,  $\kappa_0/T$ in SmB$_6$ has been controversial.  While $\kappa_0/T$ of SmB$_6$ has been reported to be very small but finite \cite{Hartstein},  the absence of  $\kappa_0/T$ has been reported in \cite{Boulanger,YXu}.  

  \begin{figure*}[t!]
	\centering
	\includegraphics[width=0.9\linewidth]{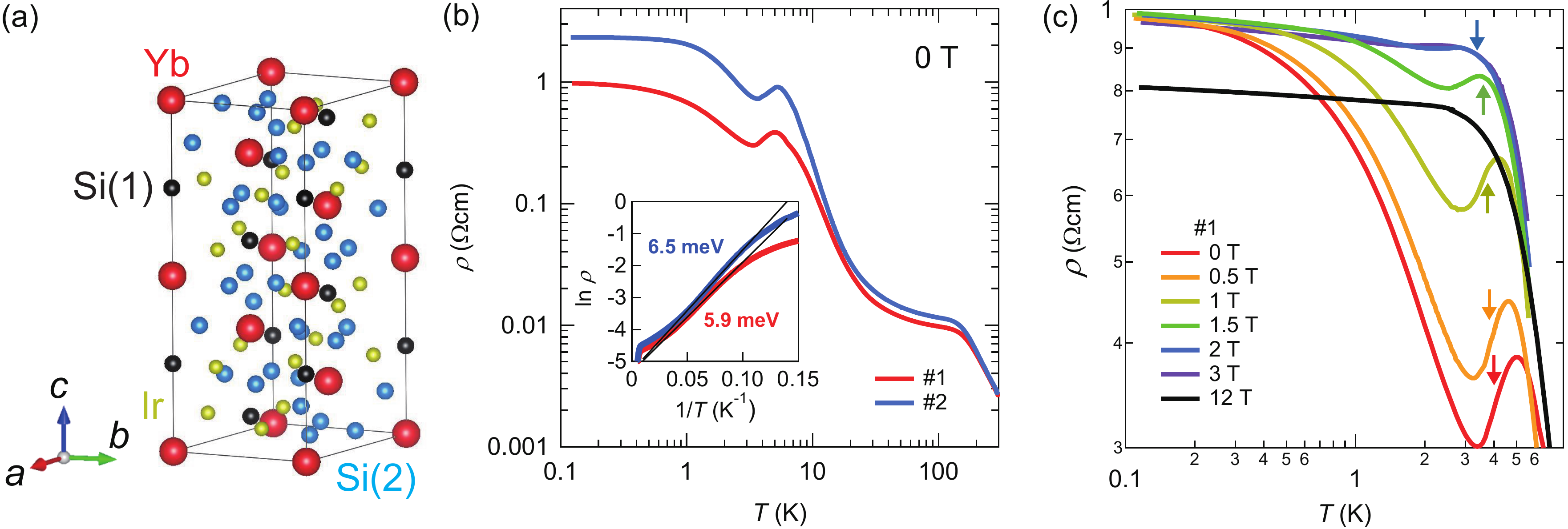}
	\caption{(a) Crystal structure of YbIr$_3$Si$_7$.  There are two crystallographically inequivalent Si sites, Si(1) and Si(2).   (b) Temperature dependence of the in-plane resistivity $\rho$ for \#1 and \#2 single crystals.  The inset  shows the Arrhenius plot of ln($\rho$) versus $1/T$.  The solid lines represent the thermally activated behaviors with charge gaps of 5.9 and 6.5\,meV for  \#1 and \#2 single crystals, respectively.  (c) Low temperature resistivity of \#1 crystal in magnetic fields applied perpendicular to the $ab$ plane.  The arrows indicate the N\'{e}el temperature determined by the specific heat.}
	\label{fig:Res}
\end{figure*}

A fascinating question is whether the QOs have any relationship to the neutral fermions.    It has been shown that $\kappa_0/T$ depends on magnetic fields in YbB$_{12}$, suggesting that the neutral fermions can couple to magnetic fields \cite{Sato}.    Various theoretical models that invoke novel itinerant low-energy neutral excitations within the charge gap that can produce QO signals have been proposed, including Majorana Fermi liquids \cite{Erten,Varma,Baskaran} and a spin liquid with spinon Fermi surface \cite{Chowdhury,Sodemann}.   
However, the nature of the neutral fermions is largely elusive  and  continues to be hotly debated.   As the Kondo hybridization between magnetic moments and conduction electrons  is the origin of the charge gap formation in these insulators, it is crucially important to clarify how the neutral fermions couple to the magnetic degrees of freedom.  Thus,  more systematic investigations on a new class of materials are highly desired to clarify the relationship between QOs, charge-neutral fermions, and magnetic properties.

 Recently a new insulating Kondo lattice compound YbIr$_3$Si$_7$ has been discovered \cite{Stavinoha}. YbIr$_3$Si$_7$ has a trigonal ScRh$_3$Si$_7$-type crystal structure (Fig.\,\ref{fig:Res}(a)).     
 The magnetization and neutron diffraction data show  that Yb-ions are very close to the trivalent state in the bulk~\cite{Stavinoha}.   In zero field,  antiferromagnetic (AFM)  order occurs below the N\'{e}el temperature $T_N=4.0$\,K.   Neutron diffraction measurements report~\cite{Stavinoha} that, in the AFM state corresponding to the $\Gamma_1$ state,  all the Yb$^{3+}$ moments are oriented along the crystallographic $c$ axis ([001]). Each Yb$^{3+}$  moment is aligned anti-parallel with its six nearest neighbors in the nearly cubic Yb sublattice and parallel with its co-planar next nearest neighbors.  The ordered moment is $\sim$1.5\,$\mu_B$/Yb$^{3+}$.   We note that  in YbIr$_3$Si$_7$,  the number of free charge carriers has been suggested to be much fewer than the number of local moments \cite{Stavinoha}.  It has therefore been proposed~\cite{Stavinoha} that the system becomes insulating at low temperatures as all the free carriers are consumed in the formation of Kondo singlets.  Thus YbIr$_3$Si$_7$  has insulating bulk and long-range magnetic correlations, and  is distinct from other simple Kondo insulators,  such as SmB$_6$ and YbB$_{12}$.  Interestingly, thickness analysis of the electric transport shows that YbIr$_3$Si$_7$ harbours conducting surface states whose origin is however not topological but rather has to do with the valence change to Yb$^{2+}$ near the sample surface~\cite{Stavinoha}.

In this paper, we investigate the low-energy excitations  in the AFM insulating state of  YbIr$_3$Si$_7$ by the low-temperature specific heat and thermal conductivity measurements.   We find that both $\gamma$  and $\kappa_0/T$ are finite at low fields, demonstrating the presence of mobile and gapless excitations of neutral fermions in the bulk insulating state, i.e.  YbIr$_3$Si$_7$ is a charge insulator but a thermal metal.   The AFM order of this compound can be widely tuned by the external  magnetic fields.  More precisely, the charge-neutral quasiparticle excitations are either gapless or gapped with an extremely small excitation energy gap, much smaller than the base temperature 90 mK of our themal conductivity measurements.
Most surprisingly, a spin-flop transition from AF-I to AF-II phase at $\mu_0H\approx$2.5\,T gives rise to an opening of a tiny gap or a linearly vanishing density of states (DOS) of neutral fermions, indicating a transition from a thermal metal into an insulator/semimetal.  These results suggest that spin degrees of freedom directly couple to the neutral fermions, whose emergent Fermi surface undergoes a transformation in applied field.
 
\section{Magnetic phases}

\subsection{Resistivity}
 Figure\,\ref{fig:Res}(b) depicts the $T$-dependence of the in-plane resistivity $\rho$ of  YbIr$_3$Si$_7$ single crystals (\#1 and \#2) plotted on a log-log scale.   At $T\sim$ 150\,K, $\rho(T)$ changes its slope, which is attributed to  the onset of Kondo correlations.   Below $\sim$150\,K, $\rho(T)$  increases rapidly  with decreasing $T$.  As   shown in the inset of Fig.\,\ref{fig:Res}(b), $\rho(T)$  increases exponentially as $\rho(T)\propto \exp{(\Delta_c/k_BT)}$ with the charge gap $\Delta_c\sim$ 5.9 and $\sim$ 6.5\,meV for sample \#1 and \#2, respectively.   At around $T_N$, $\rho(T)$ is suppressed and increases again with decreasing $T$ down to $\sim0.3$\,K.    Upon further reducing the temperature, $\rho(T)$ saturates down to the lowest temperature.   Figure\,\ref{fig:Res}(c) depicts the low-temperature resistivity in magnetic field applied parallel to the $c$ axis ({\boldmath $H$}$\parallel c$).   The suppression of $\rho(T)$ at $T_N$ is reduced in magnetic field and is absent above $\mu_0H$ = 3\,T, consistent with the previously reported  data~\cite{Stavinoha}. 

   \begin{figure}[b]
	\centering
	\includegraphics[width=0.85\linewidth]{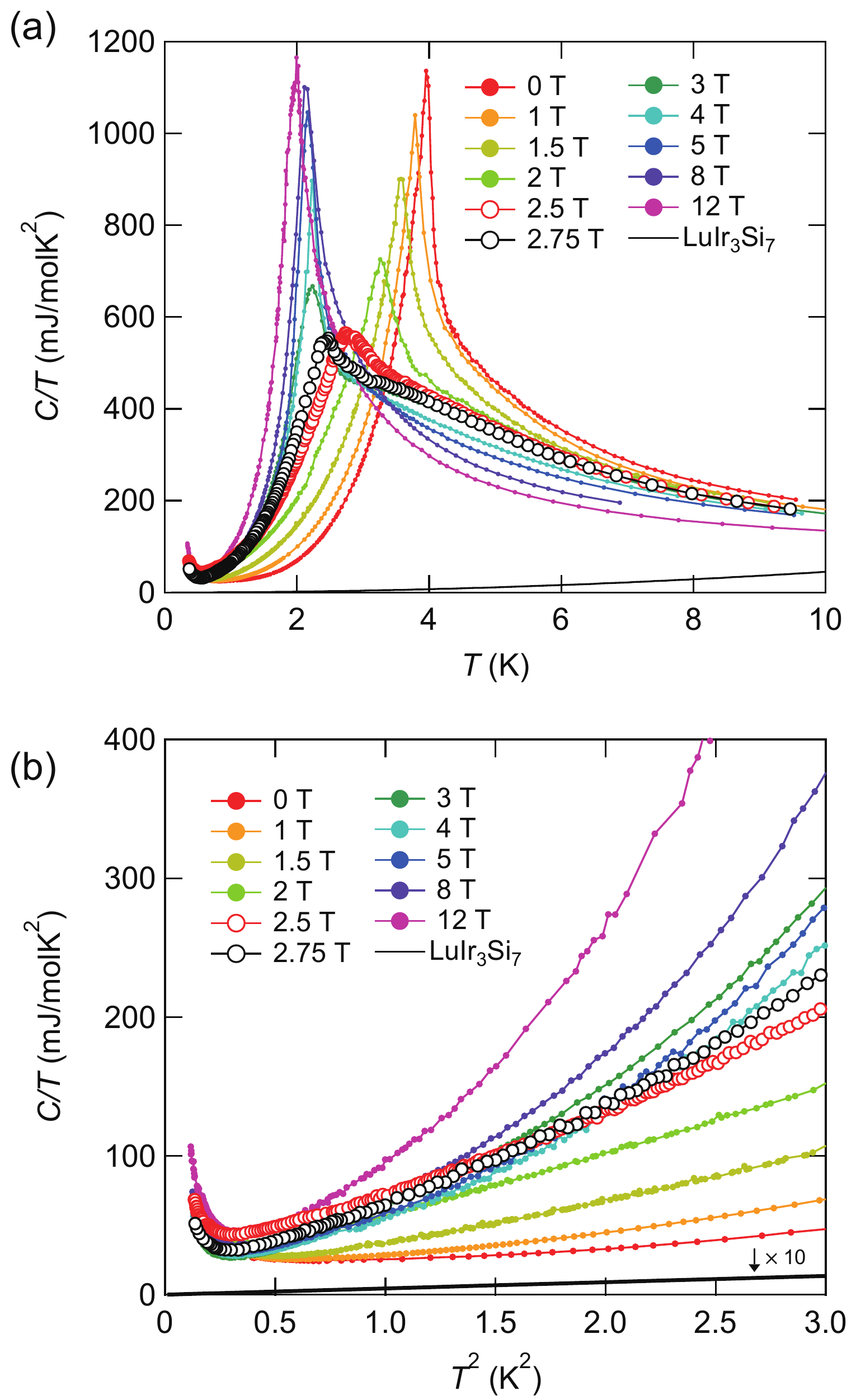}
	\caption{(a) Temperature dependence of the specific heat divided by temperature $C/T$ of YbIr$_3$Si$_7$ in magnetic field perpendicular to the $ab$ plane.  Solid line represent $C/T$ of  nonmagnetic and isostructural LuIr$_3$Si$_7$.  (b) $C/T$ vs. $T^2$ at low temperatures  in the magnetically ordered states.  
	}
	\label{fig:SH}
\end{figure}

 It has been shown that the low-temperature saturation of $\rho(T)$ arises from the surface state \cite{Stavinoha}.   In fact, the difference of the saturation values of $\rho$ between crystals \#1 ($\sim1\,\Omega$cm) and \#2 ($\sim2.1\,\Omega$cm)  can be quantitatively explained by the area and thickness of the crystal planes used for the measurements.   Similar phenomena have  been reported in  SmB$_6$ and YbB$_{12}$, in which  the metallic conductivity takes place at the surfaces of the crystal,  while electronic transport and optical measurements clearly show the opening of a finite charge-gap in the bulk at low temperatures.  These metallic surface in  SmB$_6$ and YbB$_{12}$ has been attributed to the topological insulating properties at low temperatures \cite{Dzero}.   
  In fact, the metallic surface states have been resolved by angle-resolved photoemission spectroscopy (ARPES) \cite{NXu,Hagiwara}.   In particular, spin-ARPES experiments in SmB$_6$ have revealed the spin-momentum locking of the surface quasiparticles as expected from topologically protected Dirac cones \cite{NXu}.     
 In YbIr$_3$Si$_7$, on the other hand, the recent photoemission spectroscopy measurements revealed that the surface conduction originates from a change of valence from Yb$^{+3}$ in the bulk  to Yb$^{+2}$ on the surface, without invoking topological arguments \cite{Stavinoha}. 
 
 \subsection{Phase diagram}

\begin{figure}[t]
	\centering
	\includegraphics[width=\linewidth]{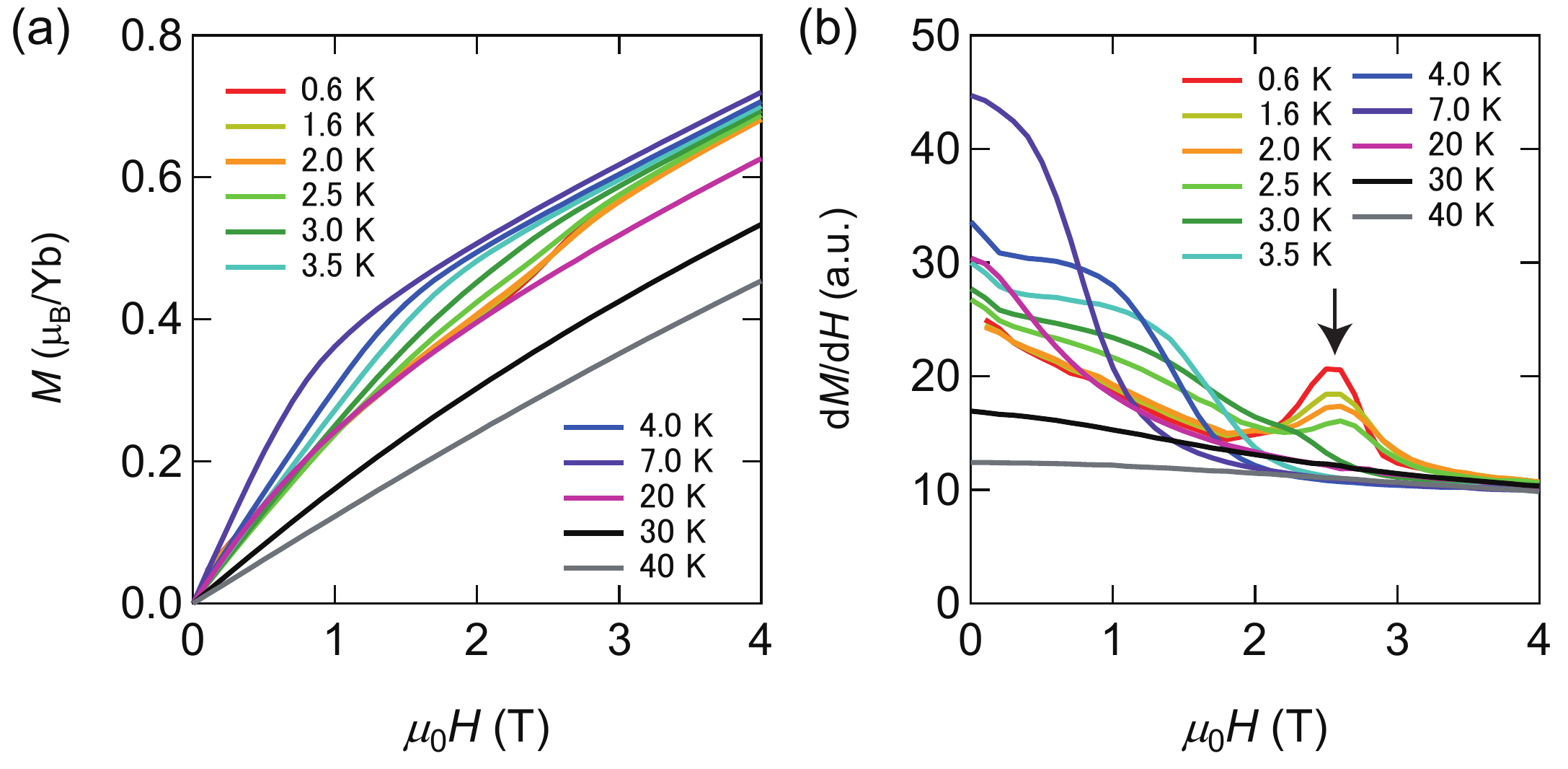}
	\caption{(a)Field dependence of the magnetization $M$ for {\boldmath $H$}$\parallel c$.  (b)Field dependence of  $dM/dH$. Peak indicated by the arrow corresponds to the boundary between AF-I and AF-II phases. 
	 }
	\label{fig:M}
\end{figure}

 Figure\,\ref{fig:SH}(a) displays the $T$-dependence of the specific heat divided by temperature, $C(T)/T$ of crystal~\#1 in zero and finite magnetic fields applied for {\boldmath $H$}$\parallel c$.   Below 10\,K,  $C(T)/T$ increases gradually with decreasing $T$  and shows a very sharp peak at $T_N$ = 4.0\,K in zero field.  As indicated by arrows in Fig.\,\ref{fig:Res}(c), the resistivity shows an anomaly at around $T_N$ determined by the specific heat.      This temperature dependence of $C/T$ is similar to that reported in metallic CeRhIn$_5$ that undergoes an AFM transition \cite{Knebel}.     The enhancement of the specific heat above $T_N$ might indicate the importance of short-range order.  In YbIr$_3$Si$_7$, the magnetic field suppresses the peak height considerably and shifts $T_N$ to lower temperatures.   The temperature dependence of $C/T$ changes dramatically at higher fields \cite{Stavinoha}.   Above $\mu_0H\approx 3$\,T,   $C(T)/T$ again exhibits a sharp peak, and the peak height increases rapidly,  followed by a nearly saturated behavior above $\mu_0H=$ 5\,T.   In contrast to lower fields, $T_N$ is nearly independent of applied magnetic field.     
 Figure\,\ref{fig:SH}(b) depicts $C/T$ plotted as a function of $T^2$ at low temperatures.      
 An upturn of $C(T)/T$ at very low temperature ($T\alt$ 0.6\,K) is attributed to the Schottky anomaly. 
 
The specific heat data clearly indicate the presence of two distinct AFM phases, i.e.,  low-field AF-I and high-field AF-II phases.   To determine the phase boundary  between these two phases  below $T_N$, we measured the $H$-dependence of the magnetization $M$ for ${\bm H}\parallel c$, as shown in Fig.\,\ref{fig:M}(a).   At around $\mu_0H\approx2.5$\,T, $M(H)$ curves show inflection points  at low temperatures.  To see this more clearly, we plot  the field derivative of the magnetization $dM/dH$ in Fig.\,\ref{fig:M}(b).  At low temperatures, $dM/dH$ shows a distinct peak as a function of $H$, which is attributed to the phase transition between AF-I and AF-II phases.   The peak field of $dM/dH$ is independent of temperature.   
In addition, no discernible hysteresis is observed between up-sweep and down-sweep magnetization measurements. Therefore, AF-I and AF-II phases are likely  separated by a weak first-order phase transition.

\begin{figure}[t]
	\centering
	\includegraphics[width=0.85\linewidth]{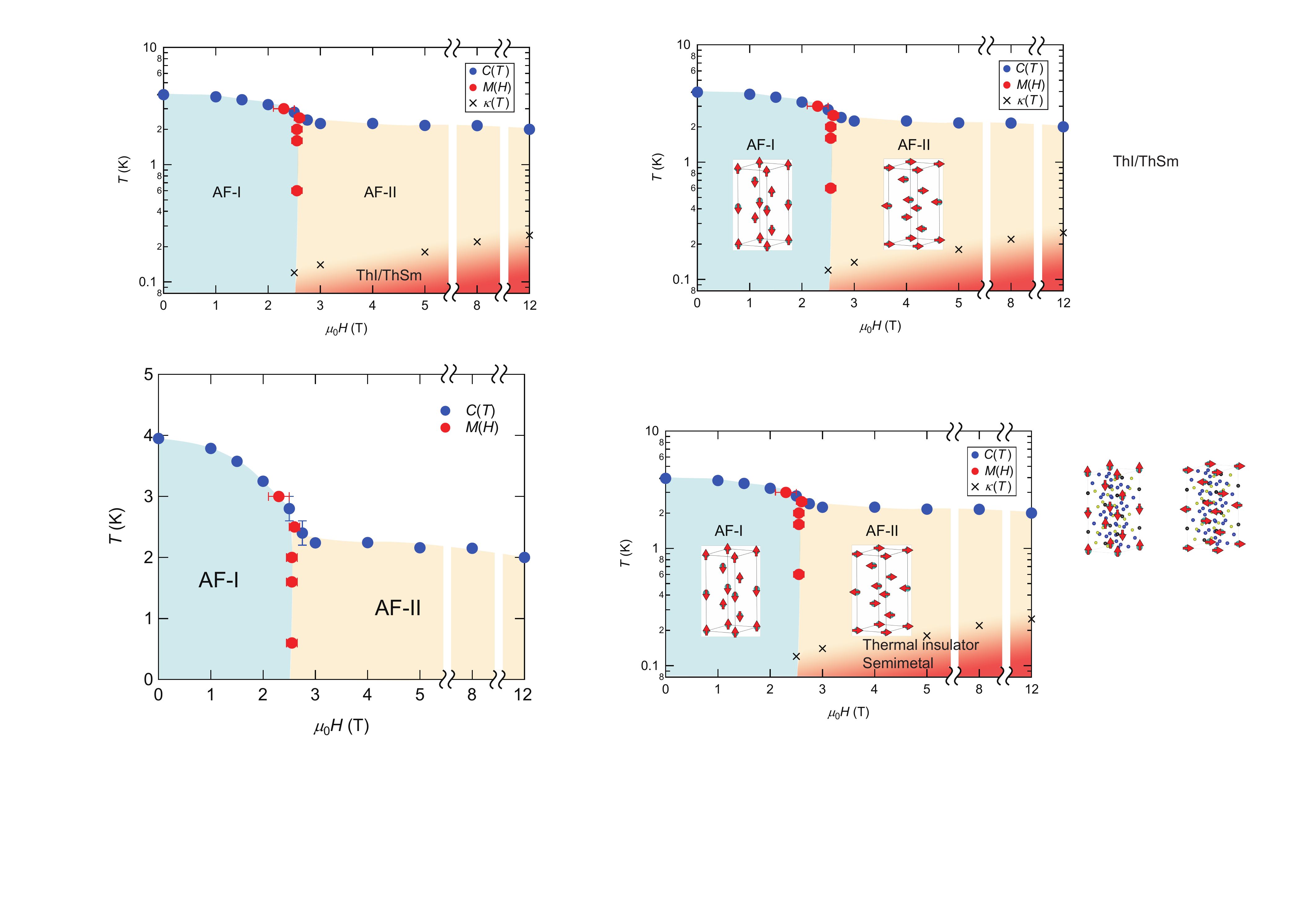}
	\caption{ $H$--$T$ phase diagram of YbIr$_3$Si$_7$ for ${\bm H}\parallel c$ axis  determined by the specific heat and magnetization. The N\'{e}el temperatures  (filled blue circles) are  determined by the peak temperature of $C(T)/T$ and phase boundary (red filled circles) is determined by the peak of $dM/dH$.   In the AF-I phase, the spins are oriented along the $c$ axis.  The AF-II phase is in the spin-flop phase, where the spins are oriented in the $ab$ plane.  The crosses represent the temperatures at which gap formation occurs, which is determined by the deviation of  $\kappa/T$ from $T^2$-dependence shown by arrows in Figs.\,\ref{fig:TC}(d)-(h).  The red colored  regime represents thermal insulator or thermal semi-metal (ThI/ThSm) regime.  }
	\label{fig:PD}
\end{figure}

\begin{figure}[t]
	\centering
	\includegraphics[width=0.8\linewidth]{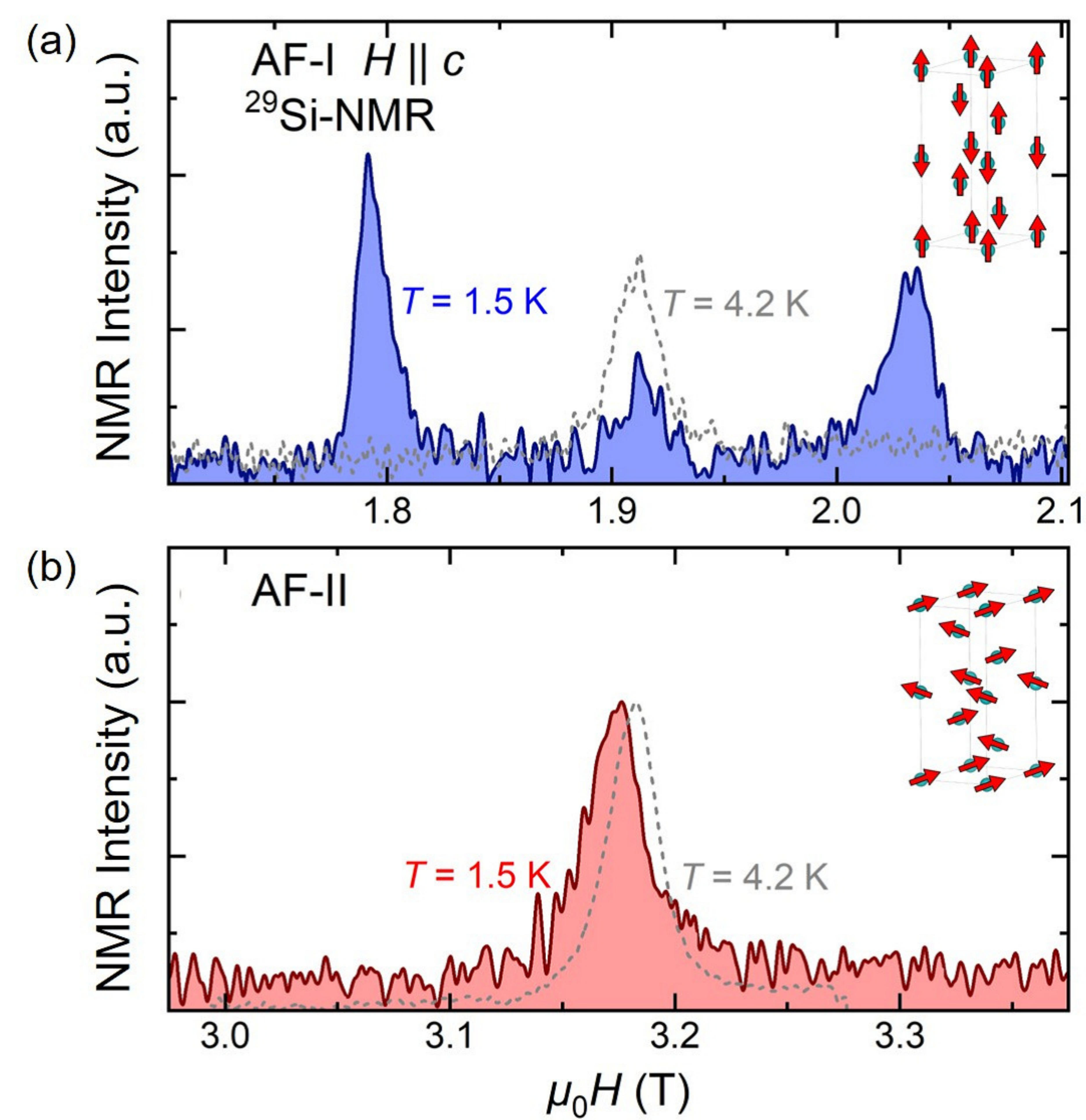}
	\caption{Magnetic-field swept $^{29}$Si-NMR spectrum (a) in the AF-I phase and  (b) in the AF-II  phase.
		Gray dotted lines represent the NMR spectrum at 4.2 K(paramagnetic state). In the AF-I phase, the NMR spectrum split into three peaks, while in the AF-II phase, only one peak is observed. The NMR results indicate the spin-flop transition.   
		The expected magnetic structure in each phase is shown in the insets.}
	\label{fig:NMR}
\end{figure}

Figure\,\ref{fig:PD} displays the $H$-$T$ phase diagram for ${\bm H}\parallel c$ axis  determined by the specific heat and magnetization measurements.   The N\'{e}el temperatures are determined by the peak temperature of $C(T)/T$. To obtain information on the nature of the phase transition, we performed nuclear magnetic resonance (NMR) measurements for {\boldmath $H$}$\parallel c$.  Figures\, \ref{fig:NMR}(a) and (b) depict the magnetic-field swept $^{29}$Si-NMR spectrum in the AF-I and AF-II phases, respectively.   There are two crystallographically inequivalent Si sites, Si(1) and Si(2), as illustrated in Fig.\,1(a).   For comparison, the NMR spectrum at 4.2\,K above $T_N$ are also shown by gray dotted lines.   In the AF-I phase, the NMR spectrum splits into three peaks. The peaks in the higher and lower magnetic fields indicate that an internal magnetic field at the Si(2) site is parallel to the external magnetic field, which is shown in the inset of Fig.\,\ref{fig:NMR}(a).   This spin structure is consistent to that reported by neutron diffraction measurements.   The middle peak arises  from the Si(1) site at which an internal magnetic field from the Yb magnetic moment is canceled.    On the other hand,  in the AF-II phase, only one peak is observed.  This peak slightly shifts to a lower field below $T_N$.  This small shift  suggests that  dominant magnetic moments are oriented perpendicular to the external magnetic field, as shown in the inset of Fig.\,\ref{fig:NMR}(b),  although the tilted angle from the $ab$ plane cannot be determined precisely in the present measurements.  Thus the NMR experiment reveals the spin-flop transition in which the magnetic moments oriented along the $c$ axis in the AF-I phase  are rotated  to the $ab$ plane in the AF-II phase.

\section{Gapless excitations in the insulating state}

\begin{figure*}[t]
	\centering
	\includegraphics[width=\linewidth]{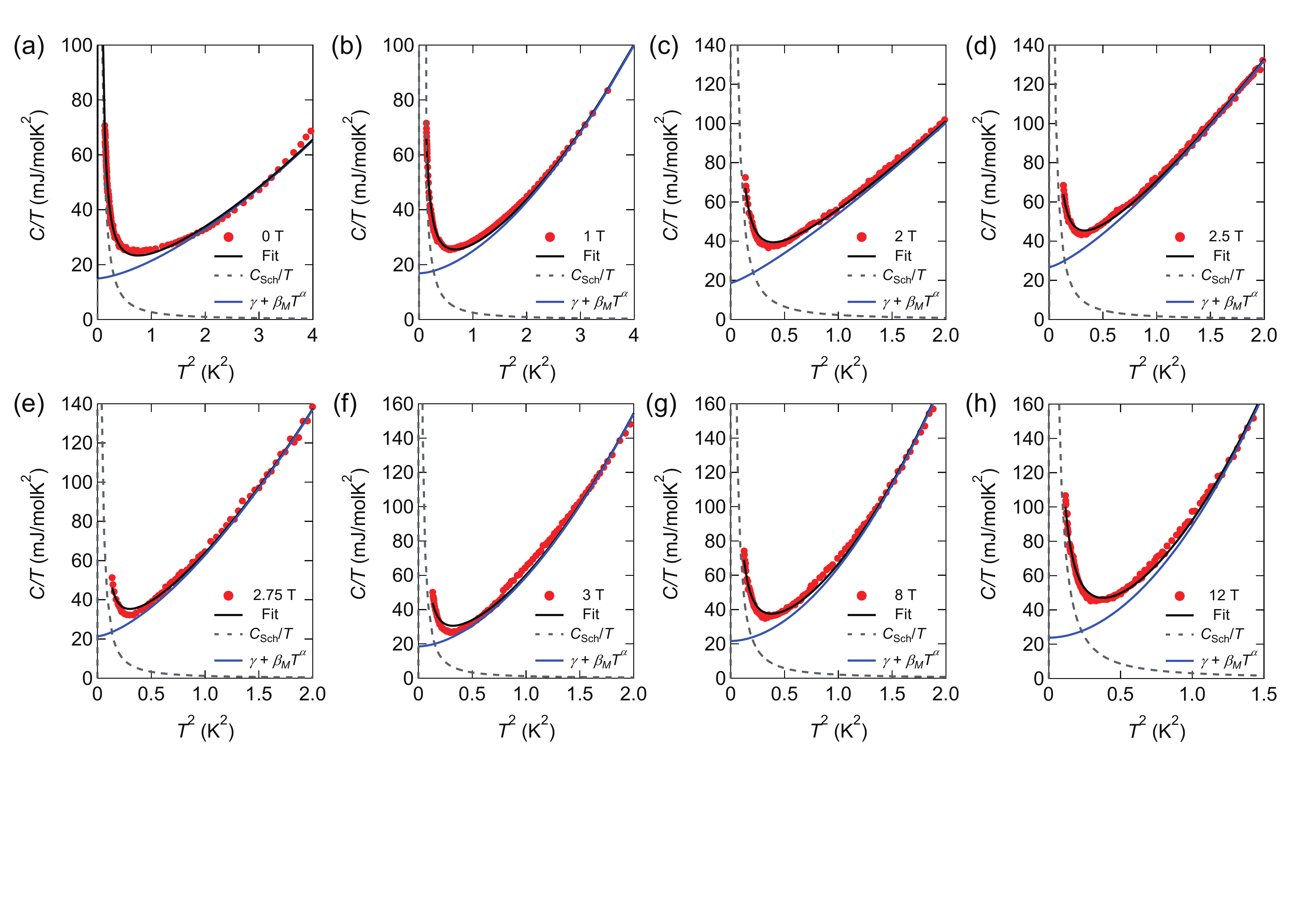}
	\caption{(a)-(h)$C/T$ vs. $T^2$ at several fields at low temperatures.    The black solid, gray dashed and blue solid lines represent the total $C/T$, Schttoky contribution, and $\gamma+\beta_M T^{\alpha}$ term, respectively, which are obtained by the fitting  using Eq.\,(1). }
	\label{fig:S2}
\end{figure*}

\subsection{Specific heat}
 The specific heat of nonmagnetic and isostructural LuIr$_3$Si$_7$ is plotted in Figs.\,\ref{fig:SH}(a) and \ref{fig:SH}(b) to estimate the phonon contribution.  The phonon specific heat is negligibly small in the whole temperature range.     
 As shown in Fig.\,\ref{fig:SH}(b), $C(T)/T$  at low temperatures varies rapidly with $T$ at high fields.   As the field is lowered, the $T$-dependence becomes weaker.    Except for the very low $T$-regime, where $C(T)/T$ shows an upturn due to  the Schottky anomaly, $C(T)/T$ increases with upward curvature with increasing $T$.

Figures\,\ref{fig:S2}(a)-(h) display $C(T)/T$ vs. $T^2$ at low temperatures.  Obviously,  the extrapolation of $C(T)/T$ above 1\,K to $T=0$  yields finite intercepts for all fields, indicating the presence of a finite $\gamma$.    As shown in these figures, the specific heat can be fitted by 
 \begin{equation}
  \frac{C(T)}{T}=\gamma+\beta_M T^{\alpha}+\frac{C_{Sch}(T)}{T}, 
  \end{equation}
for all fields.  Here, $\alpha$ is the exponent of power-law temperature dependence in the second term with the coefficient $\beta_M$, and $C_{Sch}(T)=\frac{\Delta^2}{k_BT^2}\frac{e^{\Delta/k_BT}}{(1+e^{\Delta/k_BT})^2}$  is the  two-level  nuclear  Schottky term, where $\Delta$ is the  corresponding  energy splitting.   
As seen in Fig.\,\ref{fig:SH}(b), $C/T$ increases steeper than $C/T\propto T^2$ line in the whole field range,  indicating that $\alpha$ is larger than 2.  The AFM spin-wave theory predicts $\alpha=1$ and 2 for quasi-2D and 3D systems, respectively.  Therefore, the results demonstrate the presence of contributions other than spin waves.

 \begin{figure}[b]
 	\centering
 	\includegraphics[width=0.8\linewidth]{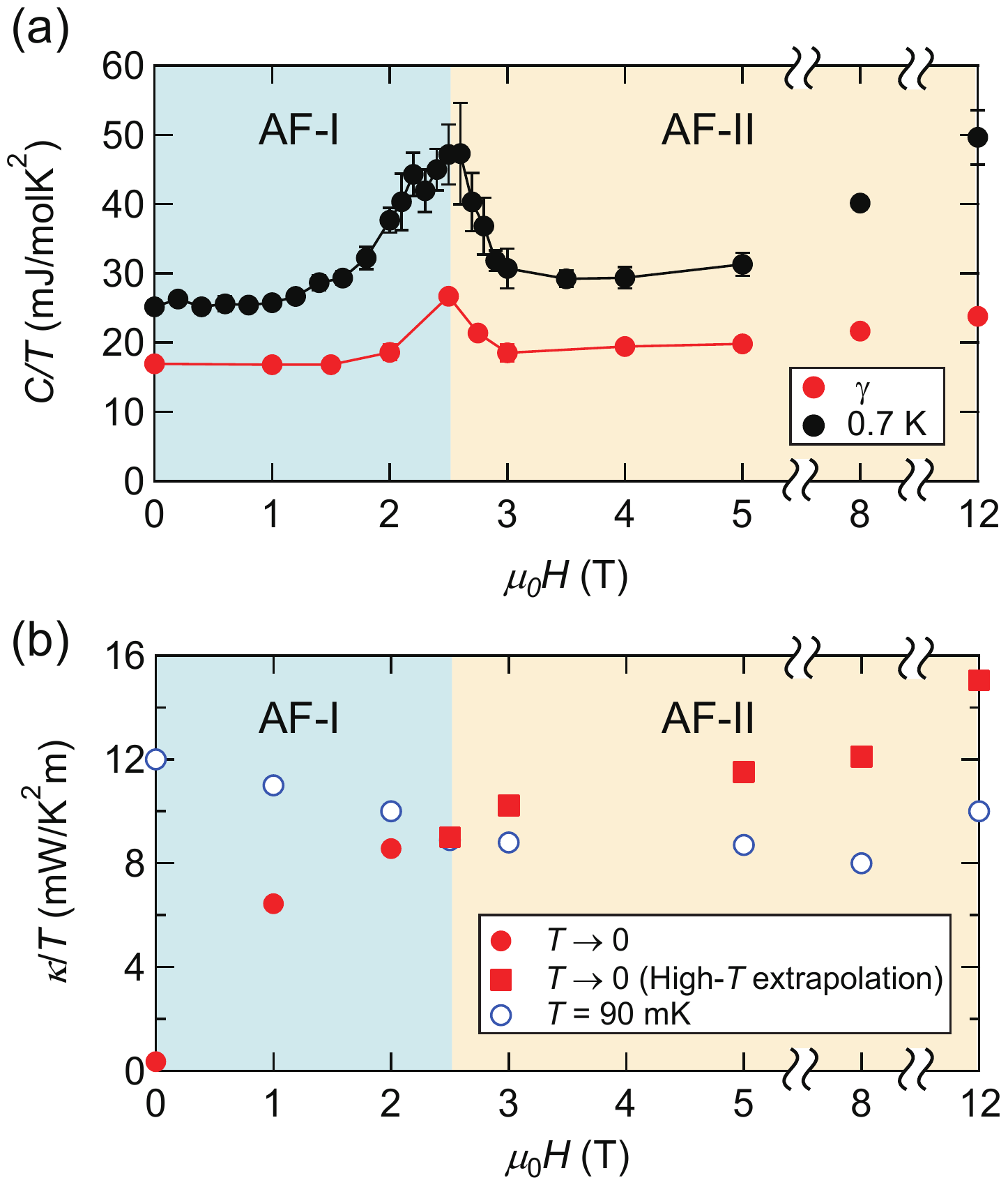}
 	\caption{(a) Field dependence of $\gamma$, which is obtained by the fitting using Eq.\,(1) (see Figs.\,\ref{fig:S2}(a)-(h)), and $C/T$ at 0.7\,K.  (b) Field dependence of the residual thermal conductivity $\kappa_0/T$ (red filled circles) in the AF-I phase and $\kappa/T$ obtained by the extrapolation from high temperature regimes to $T=0$ in the AF-II phase (red filled squares).  Open blue circles represent $\kappa/T$ at 90\,mK.  }
 	\label{fig:gamma}
 \end{figure}

 In Fig.\,\ref{fig:gamma}(a),  the $H$-dependence of $\gamma$ obtained by the fitting of Eq.\,(1) is shown.    In the whole field regime, $\gamma$ is finite.  In the low-field regime, $\gamma$ is nearly constant.  Remarkably, $\gamma$ is enhanced above $\sim$ 2\,T, and peaks in the vicinity of the phase boundary.  Upon entering the AF-II phase,  $\gamma$ is first suppressed but then increases gradually with $H$.   To confirm that this $H$-dependence of $\gamma$ is not due to a fitting ambiguity, we also plot $C(T)/T$ at 0.7\,K, where the Schottky contribution is negligible.    The similar $H$-dependence of $C(0.7\,{\rm K}) /T$ indicates that the enhancement of $\gamma$ near the phase boundary is an intrinsic property.     As the system is insulating, finite $\gamma$ indicates the presence of a finite DOS of charge-neutral excitations.  More precisely,  the charge-neutral quasiparticle excitations in the  AF-I and II phases are either gapless or gapped with an extremely small excitation energy gap, much smaller than 0.7\,K.   We shall discuss this issue in more detail below.

 \begin{figure*}[t]
 	\centering
 	\includegraphics[width=\linewidth]{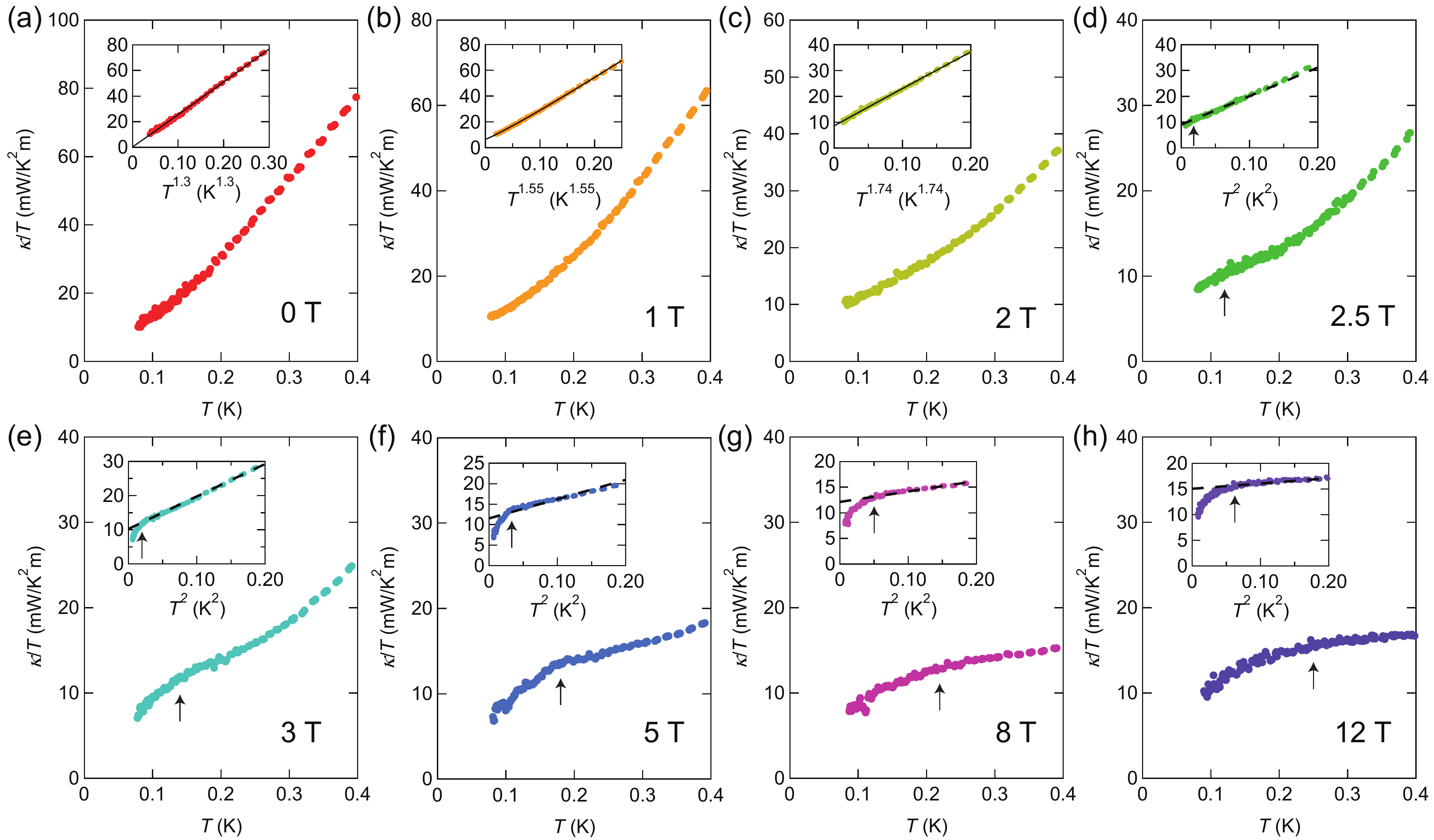}
 	\caption{(a)-(h) Thermal conductivity divided by temperature $\kappa/T$ versus  $T$ in zero and magnetic field for {\boldmath $H$}$\parallel c$  at very low temperatures.  The insets of (a), (b) and (c) show $\kappa/T$ plotted as a function of $T^p$ with $p$=1.3, 1.55 and 1.74, respectively.  The insets of (d)-(h)  show $\kappa/T$ vs. $T^2$.  The solid straight lines of (a)-(c) represent the results of the fitting. The dashed straight lines in (d)-(h) represent the linear extrapolation from the high temperature regimes.  Arrows in the main panels and insets indicate the temperatures at which $\kappa/T$ deviates from the $T^2$-dependence.}
 	\label{fig:TC}
 \end{figure*}

The strong suppression of the peak height of $C(T)/T$ and the reduction of $T_N$ with approaching the phase boundary between the  AF-I and AF-II phases suggest a possible influence from a putative  field-induced AFM quantum critical point (QCP).   In fact,  the magnitude of the magnetic moment is expected to be strongly reduced with approaching an AFM QCP,  which leads to the suppression of the peak height of the specific heat, as reported in CeRhIn$_5$ \cite{Knebel}.  In YbIr$_3$Si$_7$, however,  the putative AFM QCP is avoided by a transition into the AF-II phase.   Nevertheless, the quantum critical fluctuations  emanating from an avoided QCP in the AF-II phase would lead to the reduction of the magnetic moment.   These results lead us to consider that the enhancement of $\gamma$ in the AF-I phase near the phase boundary is caused by the AFM quantum critical fluctuations.  The striking enhancement of $\gamma$ near the AFM QCP has been reported in several classes of strongly correlated electron systems, including heavy fermions \cite{Knebel} and iron pnictides \cite{Shibauchi}.  The present results suggest that fluctuations emanating from an avoided AFM  QCP largely modify the DOS of the neutral fermions.   

 \subsection{Thermal conductivity}
 
The specific heat involves both localized and itinerant excitations.  Therefore, a finite $\gamma$ does not always indicate the presence of mobile gapless excitations. In fact, amorphous solids and spin glasses exhibit a finite $\gamma$, although the excitations in these systems are localized. Moreover, the Schottky anomaly in the specific heat often prevents the analysis of $C$ at very low temperatures.  In contrast, the thermal conductivity is determined exclusively by itinerant excitations.  In addition, it is free from the Schottky anomaly, enabling us to extend the measurements down to lower temperatures.  In particular, a finite intercept  $\kappa_0/T$  provides the most direct and compelling evidence for the presence of the itinerant and gapless fermionic excitations, analogous to the excitations near the Fermi surface in pure metals.

Figures\,\ref{fig:TC}(a)-(h) show $\kappa/T$ of crystal \#1  plotted as a function of temperature in zero and finite magnetic fields for {\boldmath $H$}$\parallel c$  at very low temperatures.   In the AF-I phase, the temperature slope of $\kappa/T$ decreases continuously with decreasing temperature.    We find that $\kappa/T$ for $\mu_0H$=0, 1, and 2\,T is well fitted as  $\kappa/T= \kappa_0/T + c_1T^p$ ($c_1$ is a constant) with $p$ = 1.3, 1.55, and 1.74,  as shown by the solid straight lines in the insets of Figs.\,\ref{fig:TC}(a), (b), and (c).   As shown in Figs.\,\ref{fig:TC}(d)-(h), the behavior of the thermal conductivity in the AF-II phase at $\mu_0H\geq2.5$\,T is fundamentally different from that in the AF-I phase; the temperature dependence of $\kappa/T$ shows a concave downward curvature below $\sim$0.3\,K.   As shown by dashed lines in the insets of Figs.\,\ref{fig:TC}(d)-(h),  $\kappa/T$ increases nearly proportional to $T^2$ in the high-temperature regime.    At very low temperatures, $\kappa/T$ deviates from the $T^2$-dependence.  

 In insulating systems, the thermal conductivity can be written as a sum of the phonon and non-phononic quasiparticle contributions, $\kappa=\kappa_{ph}+\kappa_{qp}$.   We first discuss the phonon contribution to the total thermal conductivity.   When the mean free path of acoustic phonons $\ell_{ph}$ exceeds the average size of the samples at low temperatures, the phonons undergo specular or diffuse scattering from crystal boundaries. The phonon conductivity in the boundary-limited scattering regime at low temperature is expressed as  $\kappa_{ph}=\frac{1}{3}\beta_{ph} \langle v_s\rangle \ell_{ph}T^3$, where $\beta_{ph}$ is the phonon specific heat coefficient ($C_{ph} = \beta_{ph} T^3$), and $ \langle v_s\rangle$ is the mean acoustic phonon velocity.  For diffuse scattering, $\ell_{ph}$ becomes $T$-independent, resulting in $\kappa_{ph} \propto T^3$.  On the other hand, in the case of specular reflection, $\ell_{ph}$ follows a $T^{-1}$-dependence, leading to $\kappa_{ph}\propto T^2$.  In real systems, $\kappa_{ph} \propto  T^a$  with $a$ of intermediate value between 2 and 3.    Since the temperature dependence of  $\kappa/T$ in the low temperature regime of the AF-II phase shown in Figs.\,\ref{fig:TC}(d)-(h) is expressed as $\kappa/T\propto T^q$ with $q$ less than unity, it cannot be explained by the phonon conductivity.    As shown in the inset of Fig.\,\ref{fig:TC}(h), the $T^2$ coefficient of $\kappa/T$ in the high temperature regime at $\mu_0H=12$\,T is very small, indicating negligibly small phonon thermal conductivity.     The spin-phonon scattering reduces the phonon mean free path.  As the magnetic field tends to align the spin direction, the spin-phonon scattering is reduced at high magnetic fields.    Therefore, the results of $\kappa/T$ at 12\,T provides an upper limit of the phonon contribution.     Thus these considerations indicate that the phonon contribution to the total thermal conductivity is negligibly small at low temperatures, i.e. the thermal conductivity is dominated by non-phononic quasi-particles ($\kappa\approx \kappa_{qp}$).

In zero field,  the linear extrapolation of $\kappa/T$ to $T$ = 0 has almost a zero intercept, indicating a zero or vanishingly small value of $\kappa_0/T\alt$ 1\,mW/K$^2$m.   On the other hand,   the  extrapolation to $T$ = 0 has finite intercepts  for $\mu_0H$ = 1 and 2\,T, yielding $\kappa_0/T\approx$ 6$\pm1$ and 8$\pm1$\,mW/K$^2$m, respectively.  These results demonstrate the presence of mobile and gapless fermionic excitations in the AF-I phase.   We stress that the observed finite $\kappa_0/T$  does not originate  from charged quasiparticles, in contrast to conventional metals. Evidence for this is provided by the spectacular violation of the Wiedemann-Franz (WF) law, which connects the electronic thermal conductivity $\kappa^e$  to the electrical resistivity $\rho$.  In metals at low temperatures, the ratio $L=\kappa^e\rho/T\leq L_0$ is  satisfied, where $L_0=(\pi^2/3)(k_B/e)^2=2.44\times 10^{-8} {\rm W}\Omega {\rm K}^{-2}$ is the Lorenz number.    The values of  $\kappa_0\rho_0/T$, where $\rho_0$ is the residual resistivity,  are found to be $\sim 2.6\times10^3L_0$ and $\sim3.5\times10^3L_0$ at $\mu_0H=$1 and 2\,T, respectively. Here we used $\kappa_0/T$=6.4 and 8.6\,mW/K$^2$m at $\mu_0H$=1 and 2\,T, respectively, and  $\rho_0=0.99\,\Omega$cm for both fields.     It is highly unlikely that the surface metallic region significantly violates the WF law.  In fact,  it is well known that the WF law holds in the 2D metals, even in the quantum Hall regime.  We also note that the WF expectation of $L_0/\rho_0$ from the metallic surface is less than 2.5$\times10^{-3}$\, mWK$^{-2}$m$^{-1}$,   which is by far smaller than the experimental resolution. These results lead us to conclude that the neutral fermions in the insulating bulk state are responsible for the observed finite $\kappa_0/T$.   This suggests that, as the bulk resistivity diverges as $T\rightarrow 0$, the Lorenz number for the heat-carrying quasiparticles also diverges.  Thus, the thermal conductivity and heat capacity data under magnetic fields in the AF-I phase of YbIr$_3$Si$_7$ provide evidence for the presence of highly mobile and gapless neutral fermion excitations, which has been similarly reported in YbB$_{12}$.

We note parenthetically that finite values of both $\gamma$ and $\kappa_0/T$ in the insulating states have been reported in quantum-spin-liquid candidates with 2D triangular lattices,  including the organic compounds,  EtMe$_3$Sb[Pd(dmit)$_2$]$_2$ \cite{Yamashita} and $\kappa$-H$_3$(Cat-EDT-TTF)$_2$ \cite{Shimozawa},  and inorganic compounds,  1$T$-TaS$_2$ \cite{Murayama} and Na$_2$BaCo(PO$_4$)$_2$ \cite{Li2018}.    In EtMe$_3$Sb[Pd(dmit)$_2$]$_2$,  although the presence or absence of finite $\kappa_0/T$ has been controversial among different research groups \cite{Hope,Ni}, it has been shown very recently that differences between data sets are likely to be due to the cooling rate \cite{Yamashita2}.  In 1$T$-TaS$_2$, as finite $\kappa_0/T$ readily disappears by the introduction of disorder/impurity, the magnitude of $\kappa_0/T$ appears to depend strongly on the sample quality \cite{Murayama}.  These results suggest that high-quality single crystals are required to observe the finite $\kappa_0/T$ in quantum spin liquid systems.    In the above compounds, finite $\gamma$ and $\kappa_0/T$ have been discussed in terms of electrically neutral spinons forming the Fermi surface.
  
In the AF-II phase of YbIr$_3$Si$_7$, as shown by the dashed lines in the insets of Figs.\,\ref{fig:TC}(e)-(h),   the linear extrapolation of $\kappa/T$ from the high-temperature regime has finite intercepts at $T$ = 0.    However, remarkable deviations from the $T^2$-dependence and  suppression of $\kappa/T$ at very low temperatures  are clearly observed,  likely  indicating an opening of a tiny gap in the spectrum of the itinerant excitations.  As this gap formation occurs below $\sim$ 0.3\,K,   the estimate of the gap is two orders of magnitude smaller than the Kondo gap ($\sim 60$\,K).   We point out that there are two possible  explanations for this behavior at very low temperatures well below $\sim$ 0.3\,K.  One is a fully gapped thermal insulating state and the other is thermal semimetallic or nodal metallic state with a  linearly vanishing  DOS, as indicated by red shaded regime in Fig.\,\ref{fig:PD}.  To clarify which scenario is realized,  future  measurements at lower temperature are required.   

We note that the specific heat measurements cannot resolve this gap formation due to the Schottky anomaly.     The red filled circles in Fig.\,\ref{fig:gamma}(b) represent $\kappa_0/T$ in the AF-I phase.    After the initial rapid increase,  $\kappa_0/T$ increases slowly.    In Fig.\,\ref{fig:gamma}(b), $\kappa/T$($T\rightarrow 0$) in the AF-II phase, which is obtained by the linear extrapolation from the high-temperature regime to $T$=0,  is plotted by filled red squares.   For comparison,  we also plot $\kappa/T$ at 90\,mK.   Interestingly, $\kappa/T$ obtained by high-temperature extrapolation appears to lie on top of the extrapolation from the AF-I phase.    This suggests that  $\kappa_0/T$ steadily increases with magnetic field, but is strongly  affected by the spin-flop transition.   This field dependence indicates that the itinerant neutral fermions couple to the magnetic field and are strongly influenced by the magnetic ordering.
 
 \section{Discussion}
 
The combined results of the specific heat and thermal conductivity provide pivotal information on the neutral fermions observed in insulating materials.    As shown by Figs.\,\ref{fig:gamma}(a) and \ref{fig:gamma}(b),  $\gamma$ and $\kappa_0/T$ exhibit very different $H$-dependence.    In particular, at zero field,  while  $\gamma$ is finite,  no sizable  $\kappa_0/T$ is observed.    We note that the result at zero field bears a resemblance to that of SmB$_6$.   
On the other hand, finite  $\gamma$ and $\kappa_0/T$ values in YbB$_{12}$ are similar to those of YbIr$_3$Si$_7$ in a finite field in the AF-I phase. 
In YbB$_{12}$, the $\gamma$ value is nearly sample independent, while $\kappa_0/T$ values are strongly sample dependent, which is attributed to the amount of the impurities/defects determining the mean free path of the quasiparticles.  In contrast, in the present study, we find strong field dependence of $\kappa_0/T$ in a single sample, which cannot be due to the change in the impurity scattering.  
The rapid enhancement of  $\kappa_0/T$ at low $H$ is attributed to the increase of the number of mobile heat carriers and/or the change of the dispersion of the neutral particles, which results in the increase of the group velocity.     
The absence of an enhancement of $\kappa_0/T$ near the avoided QCP, despite the enhancement of $\gamma$, may be because $\kappa_0/T$ is proportional to $\gamma \tau$, where $\tau$ is the scattering time.  To explain why $\kappa_0/T$ is not seriously   affected  by the enhancement of $\gamma$ near the QCP, it is required that  $\tau$ is inversely proportional to the DOS of the neutral fermions, $\tau\propto 1/\gamma$.  Such a mechanism is, for example,  observed in $d$-wave superconducting materials, which show an universal residual thermal conductivity \cite{Balatsky}.

 The nature and behavior of the novel charge neutral fermions are not well understood; there are very few experimental results that can be used as tests of the various  theoretical  models, which include 3D Majorana fermions, composite magnetoexcitons, and spinons in fractionalized Fermi liquids.  In this sense, our observations that the itinerant neutral fermions are very sensitive to the magnetic ordering can put significant restrictions on the various theories.  The tiny gap formation (or a linearly vanishing DOS) in the AF-II phase indicates a  transition from a thermal metal into an insulator (or a thermal semimetal), while the material remains an electrical insulator.  This result demonstrates that the Fermi  surface of the charge-neutral fermions becomes unstable   towards gap formation at low temperatures, which is driven by the magnetic transition of the insulator.  Therefore it is natural to consider that the neutral fermions are composed of strongly magnetically coupled $c$ and $f$ electrons through the Kondo effect. In this situation, neutral fermion excitations will be affected by AFM order and fluctuations.  

 \section{Conclusions}
 
In summary, we performed specific heat, thermal conductivity and NMR measurements of bulk insulating YbIr$_3$Si$_7$ at low temperatures. 
In the low-field AF-I phase, we find finite $\gamma$ and $\kappa_0/T$,  demonstrating the emergence of itinerant gapless excitations.   A spectacular violation of the Wiedemann-Franz law directly indicates that YbIr$_3$Si$_7$ is a charge insulator but a thermal metal.  More precisely, the charge-neutral quasiparticle excitations are either gapless or gapped with an extremely small excitation energy gap, much smaller than the base temperature 90 mK of our themal conductivity measurements.  
A spin-flop transition is revealed by NMR measurements.    With approaching the spin-flop transition,  $\gamma$ is largely enhanced.  
Remarkably, inside the high-field AF-II phase,  $\kappa/T$  exhibits a  sharp drop at very low temperatures, indicating the opening of a  tiny gap much smaller than the Kondo gap or a linear vanishing DOS of the neutral excitations.   This demonstrates a field-induced  transition from a thermal metal into an insulator/semimetal  driven by the spin-flop  transition.   The present results demonstrate that the neutral fermions are directly coupled to the spin degrees of freedom. 
Moreover, the Fermi surface of the neutral fermions has an instability towards a novel gapped  or semimetallic state.
Our experimental observations impose a strong constraint on the theories of charge-neutral fermions.  
Thus  YbIr$_3$Si$_7$ provides an intriguing platform for studying the neutral fermions 
 in strongly correlated insulators.

\section*{ACKNOWLEDGMENTS}
A.H.N. and E.M. acknowledge fruitful discussions with Chris Hooley. Y.M.  acknowledges discussion with H. Kontani, Lu Li, and J. Singleton.  A.H.N. was supported by the National Science Foundation grant No. DMR-1917511 and the Robert A. Welch Foundation grant C-1818. This work is supported by Grants-in-Aid for Scientific Research (KAKENHI) (Nos. JP15H02106, JP18H01177, JP18H01178, JP18H01180, JP18H05227, JP19H00649, JP20H02600, JP18K03511 and JP20H05159) and on Innovative Areas ``Quantum Liquid Crystals" (No. JP19H05824) from Japan Society for the Promotion of Science (JSPS), and JST CREST (JPMJCR19T5). GC acknowledges the support from NSF via grant DMR
1903888.





\end{document}